\documentclass[preprint,preprintnumbers,amsmath,amssymb]{revtex4}
\usepackage[abs]{overpic}
\usepackage{graphicx,axodraw,subfigure}
\begin{document}
\preprint{HUPD1202}
\def\tbr{\textcolor{red}}
\def\tcr{\textcolor{red}}
\def\ov{\overline}
\def\nn{\nonumber}
\def\f{\frac}
\def\beq{\begin{equation}}
\def\eeq{\end{equation}}
\def\bea{\begin{eqnarray}}
\def\eea{\end{eqnarray}}
\def\bsub{\begin{subequations}}
\def\esub{\end{subequations}}
\def\dc{\stackrel{\leftrightarrow}{\partial}}
\def\ynu{y_{\nu}}
\def\ydu{y_{\triangle}}
\def\ynut{{y_{\nu}}^T}
\def\ynuv{y_{\nu}\frac{v}{\sqrt{2}}}
\def\ynuvt{{\ynut}\frac{v}{\sqrt{2}}}
\def\d{\partial}
\title{CP violation of Extended Higgs sector and
Its impact on $D^0 \to \mu^+ \mu^-$ decay} 
\author{D. Kimura,$^1$ Kang Young Lee$^2$ and 
T. Morozumi$^3$}
\affiliation{$^1$Faculty of Education, Hiroshima University,
Higashi-Hiroshima, 739-8524, Japan \\
$^2$Division of Quantum Phases and Devices, School
of Physics, Konkuk University, Seoul, 143-701, Korea \\
$^3$Graduate School of Science, Hiroshima University,
Higashi-Hiroshima, 739-8526, Japan
}
\def\nn{\nonumber}
\def\beq{\begin{equation}}
\def\eeq{\end{equation}}
\def\bei{\begin{itemize}}
\def\eei{\end{itemize}}
\def\bea{\begin{eqnarray}}
\def\eea{\end{eqnarray}}
\def\ynu{y_{\nu}}
\def\ydu{y_{\triangle}}
\def\ynut{{y_{\nu}}^T}
\def\ynuv{y_{\nu}\frac{v}{\sqrt{2}}}
\def\ynuvt{{\ynut}\frac{v}{\sqrt{2}}}
\def\s{\partial \hspace{-.47em}/}
\def\ad{\overleftrightarrow{\partial}}
\def\ss{s \hspace{-.47em}/}
\def\pp{p \hspace{-.47em}/}
\def\bos{\boldsymbol}
\begin{abstract}
We study the impact of the CP violation of  the extra Higgs sector on $D^0$
decay. The CP even and CP odd neutral Higgs mixing of the 
two Higgs doublet model is studied and we show how the
CP violating effect of the mixing
may lead to the longitudinal muon polarization asymmetry of $D^0 \to \mu^+ \mu^-$. The asymmetry of the short-distance contribution is sensitive to the
CP violating phase of the extended Higgs sector.
\end{abstract}
\maketitle
 Higgs sector is currently investigated by LHC. In this talk,
we study the extended Higgs sector which may solve one of the questions
of the origin of the flavor. 
In the standard model, all the masses of the quarks and leptons
come from a single vacuum expectation value of Higgs, $m_q=y_q v,
m_l=y_l v$.
In the extended Higgs sector with two 
Higgs doublets, the mass hierarchy of the 3rd generation comes 
from the ratio of the two vacuum expectation values of the
two Higgs doublets.  
In Ref.\cite{Hashimoto:2004kz}, the origin of the large weak isospin breaking
is explained in the context of the two Higgs doublet model.
In this case, the extra Higgs 
which couples with the  bottom quark and tau lepton has
a small vacuum expectation value.
The strength of the Yukawa coupling of the Higgs to
down type quarks and charged leptons is much larger than the corresponding couplings
of the standard model Higgs. One may probe the large 
strength of the Yukawa couplings of the extra Higgs 
by using the bottom quark and $\tau$ lepton system. 
\section{Two Higgs doublet model with large $\tan \beta$}
In the scenario, the origin of the mass hierarchy of the bottom quark, top quark, and $\tau$ lepton is the ratio of the vacuum expectation values
of the two Higgs doublets. In contrast to the standard model where
the hierarchical
Yukawa couplings to bottom and top quarks are assumed,  
the Yukawa coupling of the bottom quark to the extra Higgs is
the same order as that of the top quark in the scenario \cite{Hashimoto:2004kz}.
With the large ratio of the vacuum expectation values
$\tan \beta=\frac{v_2}{v_1} \simeq 40$, the ratio of the Yukawa couplings
becomes,
\bea
\frac{y_b v_1}{y_t v_2}=
\frac{m_b}{m_t} \rightarrow \frac{y_b}{y_t}=\frac{m_b}{m_t}\tan \beta =1. 
\eea
The strength of Yukawa couplings
of  
the extra Higgs ($\Phi_1$) to bottom quark and
$\tau$ lepton are much larger than that of the standard model Higgs,
\bea
y_b=3 y_\tau \simeq y_{t~\rm{SM}}. 
\eea
The large strength of Yukawa couplings implies the bottom quark and tau lepton probe the extended Higgs sector. One can write the Yukawa couplings explicitly as\cite{Kimura:2012nx},
\bea
{\cal L}_Y&=&\frac{H_2}{v}\Biggl{[} \tan \beta (\overline{e_i} i\gamma_5 e^{-i \gamma_5 \theta_{AH}} m_{li} e_i
+\overline{d_i} i \gamma_5 e^{-i \gamma_5 \theta_{AH}} m_{di} d_i)
                                                \Biggr{]} \nn \\
           &+&\frac{H_3}{v}\Biggl{[} \tan \beta (\overline{e_i} e^{-i \gamma_5 \theta_{AH}} m_{li} e_i
           +\overline{d_i} e^{-i \gamma_5 \theta_{AH}} m_{di} d_i) \Biggr{]}.
\eea
One can see $tan \beta$ enhancement of the Yukawa couplings.
\subsection{Higgs potential leading to large $\tan \beta$ with CP violation}
In the two Higgs doublet model with the softly broken $Z_2$ symmetry,
the Higgs potential can naturally 
lead to the large $\tan \beta$,
as shown in \cite{Hashimoto:2004kz}. There the smaller Higgs 
vacuum expectation value is proportional to
the mass parameter of the softly broken $Z_2$ symmetry,
\bea
\Phi_1 \rightarrow -\Phi_1, \Phi_2 \rightarrow \Phi_2.  \nn 
\eea
The source of CP violation of the Higgs sector is a quartic 
coupling and the Higgs potential is given in \cite{Kimura:2012nx},
\bea
V_{\rm tree}&=&\sum_{i=1,2}\left( m_{i}^2 \Phi_i^\dagger \Phi_i+
\frac{\lambda_i}{2} (\Phi_i^\dagger \Phi_i)^2 \right)
-m_{3}^2 (\Phi_1^\dagger \Phi_2 + h.c.)\nn \\
&+&\lambda_3  (\Phi_1^\dagger \Phi_1)(\Phi_2^\dagger \Phi_2)
+\lambda_4 |\Phi_1^\dagger \Phi_2|^2  \nn \\
&+& \frac{1}{2} \lambda_5 [e^{i \theta_5}(\Phi_2^\dagger \Phi_1)^2 + e^{-i \theta_5}(\Phi_1^\dagger \Phi_2)^2].\nn
\label{eq:tree}
\eea
where $m_3^2$ is the soft breaking parameter for $Z_2$ symmetry. 
The origin of the
CP violation of the vacuum expectation value of Higgs is a CP violating
phase $\theta_5$ of the Higgs potential.
The vacuum expectation values of the Higgs can be parameterized by
the three order parameters.
\bea
\langle \Phi_1 \rangle =\frac{v}{\sqrt{2}}
\begin{pmatrix}
0 \\
\cos \beta
\end{pmatrix}, \quad 
\langle \Phi_2 \rangle =\frac{v}{\sqrt{2}} \begin{pmatrix} 
0 \\
\sin \beta  \end{pmatrix} 
e^{-i \theta^\prime}.
\eea
When $\tan \beta$ is very large, the Higgs fields can be written as,
\bea
\Phi_1&=&-\sin \beta \begin{pmatrix}  H^+ \\
\frac{H+iA}{\sqrt{2}} \end{pmatrix}+
\cos \beta \begin{pmatrix} 0 \\
\frac{v+h}{\sqrt{2}} \end{pmatrix}
, \nn \\
\tilde{\Phi}_2&=& e^{i \theta^\prime}\left(
\sin \beta \begin{pmatrix}
 \frac{v+h}{\sqrt{2}} \\
0 \end{pmatrix}+\cos \beta
\begin{pmatrix}
\frac{H-iA}{\sqrt{2}}\\
 -H^-
\end{pmatrix} \right),\nn
\eea
where CP eigenstates of the neutral Higgs ($H,A$) are not mass eigenstates
and are related to the mass eigenstates  $(H_2,H_3)$ as, 
\bea
H=\cos \theta_{AH} H_3 + \sin \theta_{AH} H_2, \nn \\
A=\cos \theta_{AH} H_2-\sin \theta_{AH} H_3.
\eea
The mixing angle $(\theta_{AH})$ of CP even (H) and CP odd (A) Higgs bosons
becomes,
\bea
\theta_{AH}&=&\theta_5+\frac{\theta^\prime}{2},\nn \\
           &=& \arctan \Bigl{[}\frac{M_{H_3}^2}{M_{H_2}^2} \tan \frac{\theta_5}{2}\Bigr{]}.
\eea
For the fixed ratio of the masses of the two eigenstates, 
the mixing angle $\theta_{AH}$
can be large. For details, see figure.12 of \cite{Kimura:2012nx}.
\section{Impact of CP violation of the neutral Higgs sector
on Rare $D^0$ decay}
\begin{figure}
\includegraphics[width=7cm]{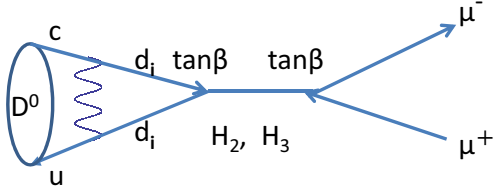}
\caption{Feynman diagram with $\tan^2 \beta $ enhancement due to
the neutral Higgs exchange in  $D^0 \to \mu^+ \mu^-$ amplitude. The wavy
internal line denotes $W^+$ boson and $H^+$. }
\label{fig:fig1}
\end{figure}

In this section, we argue how the CP violation of the
neutral Higgs boson may have some impact on a $D^0$ decay.
At first sight, the impact of CP violation of the neutral
Higgs on $D^0$ decays is small and is not relevant, since
the neutral Higgs bosons with the small vacuum expectation value is coupled to 
the down quark and 
charged lepton.  The coupling to the up quark sector is suppressed. 
However in $D^0$ FCNC
decay such as $D^0 \to \mu^+ \mu^-$, the situation is different from the 
expectation.
In the tree level Feynman diagram, the FCNC $D^0$ decay is forbidden, 
since this model respects condition for the natural 
flavor conservation. Therefore the decay occurs through one loop diagrams
where the down type quarks in the loop diagrams contribute to the process
, as shown in Fig.\ref{fig:fig1}.
Since the down quarks 
and the extra neutral Higgs couplings are large, 
one may expect that there is 
large contribution due to the new physics interaction.

The decay 
amplitudes of $D^0 \to \mu^+ \mu^-$ can be generically parameterized as,
\bea
M(D^0 \to \mu^+ \mu^-) \sim (a \bar{u} \gamma_5 v + i 
b \bar{u} v). \nn
\eea
The longitudinal muon polarization asymmetry can be written as,
\bea
P_L&=&\frac{2{\rm Im} (a b^\ast) \sqrt{1-\frac{4{m_\mu}^2}{M_D^2}}}{|a|^2+
|b|^2(1-\frac{4 {m_\mu}^2}{M_D^2})}.
\eea
In the present model, the amplitudes $a$ and $b$ are given as,
\bea  
a&=&\frac{m_{b}^2}{4 s_W^2 M_W^2}
m_\mu \Biggr{[}1+\Bigr{|}\frac{V_{cs}V_{us} m_s^2}
{V_{cb}V_{ub} m_b^2}\Bigl{|}e^{i\phi_3}\nn \\
&-& i \frac{r}{2}\Bigl{\{}
\left(\frac{M_W^2}{M_2^2}
+ \frac{M_W^2}{M_3^2}\right) +e^{i 2 \theta_{AH}} 
\left(\frac{M_W^2}{M_2^2}-\frac{M_W^2}{M_3^2}\right) \Bigr{\}}
 \Biggr{]} \nn \\
b&=&-\frac{m_{b}^2}{4 s_W^2 M_W^2}m_\mu \frac{r}{2}
\Biggl{[}\left(\frac{M_W^2}{M_3^2}+ 
\frac{M_W^2}{M_2^2}\right)
-e^{i 2 \theta_{AH}} 
\left(\frac{M_W^2}{M_2^2}-\frac{M_W^2}{M_3^2}\right) \Biggr{]},
\eea
where
$r=\frac{M_D^2 \tan^2 \beta}{M_W^2} \log \frac{M^2_{H^+}}{M^2_W}$.
The terms proportional to $r$ are new physics contribution.  
In the standard model, only a pseudoscalar term denoted by the term proportional to $a$ is generated and
it leads to vanishing asymmetry for $P_L$. 
In the present model, through the short distance
contribution due to the neutral Higgs exchange diagrams, 
there appears the large longitudinal polarization asymmetry $P_L$.
If we ignore the contribution from the standard model, the asymmetry
is approximately given as,
\bea
P_L &\simeq& -\log \frac{M^2_{H^{+}}}{M_W^2} 
\tan^2 \beta 
\sin 2 \theta_{AH} (\frac{M_D^2}{M_2^2}-\frac{M_D^2}{M_3^2}).
\eea 
The asymmetry is proportional to the neutral Higgs mixing $\theta_{AH}$.
Therefore by measuring the asymmetry, one may obtain 
the CP even and CP odd mixing angle.
In contrast to the case discussed, when there is no mixing between CP
even (H) and CP odd Higgs (A), then the CP odd Higgs 
generates the pseudoscalar term $a$ and CP even Higgs 
generates the scalar term $b$. In this case, the asymmetry vanishes.
If two neutral Higgs bosons are mixed in their mass eigenstates, a and b depend on the mixing angle of neutral Higgs and the
muon polarization asymmetry $P_L$ is sensitive to the CP violating mixing angle of Higgs.
\begin{figure}
\includegraphics[width=7cm]{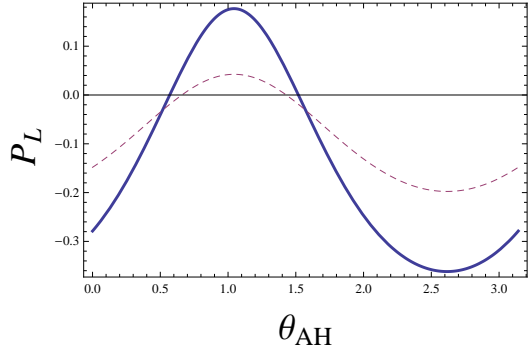}
\caption{ The muon longitudinal asymmetry for $D^0 \to \mu^+ \mu^-$
of the two Higgs soublet model. The asymmetry is shown as a function of 
CP odd and CP even Higgs mixing angle $\theta_{AH}$.
We assume
the charged Higgs mass is $500$ (GeV). We show the case for 
$M_{H_2}> M_{H_3}$ and we take,
$(M_{H_2},M_{H_3})=(200,400)$ (solidline) and $(300,600)$(dashed line).}
\label{fig:fig2}
\end{figure}
In Fig.\ref{fig:fig2}, we showed the asymmetry $P_L$ as a function of 
the micing angle of CP even and CP odd states.
\section{Conclusion and Discussion}
We study the CP violation of the two Higgs doublet model.
CP violation of the Higgs potential leads to the mixings of CP even and CP odd Higgs. The Yukawa couplings of down quarks and charged leptons with
the neutral Higgs of the second doublet are large and are CP violating.
The longitudinal muon polarizability $P_L$ of $D^0 \to \mu^+ \mu^-$
is sensitive to CP even and CP odd Higgs mixing of the two Higgs doublet
model with large $\tan \beta$. We have estimated the short-distance part of
the new physics effect. The additional contribution together with the standard
model Z penguin contribution leads to the non-zero asymmetry $P_L$. 
For more serious estimate, 
we must include the long-distance contribution from the amplitude
$D^0 \to \gamma^{\ast}
\gamma^\ast \to \mu^+ \mu^-$. The estimate of the branching fraction
from this amplitude is studied in \cite{Burdman:2001tf} and it can dominate
over the short distance contribution.
\section{Acknowledgement}
We would like to thank the organizers of GUT2012, especially to  
Dr. H. Sugiyama and Prof. T. Fukuyama.
K.Y.L. was supported by WCU program through the KOSEF funded by the 
MEST (R31-2008-000-10057-0) and the Basic Science Research Program through 
the NRF funded by MEST(2010-0010916). 
The work of T. M. is supported by KAKENHI, Grant-in-Aid for
Scientific Research(C) No.22540283 from JSPS, Japan.


\begin{thebibliography}{9}
\bibitem{Hashimoto:2004kz}
M.~Hashimoto and S.~Kanemura,
Phys.\ Rev.\ D {\bf 70}, 055006 (2004)  [Erratum-ibid.\ D {\bf 70}, 119901 (2004)]  [hep-ph/0408313]. 
\bibitem{Kimura:2012nx} 
D.~Kimura, K.~Y.~Lee and T.~Morozumi,
The Form factors of $\tau \to K \pi(\eta) \nu$ and the predictions for CP violation beyond the standard model, arXiv:1201.1794 [hep-ph].

\bibitem{Burdman:2001tf} 
  G.~Burdman, E.~Golowich, J.~L.~Hewett and S.~Pakvasa,
  Rare charm decays in the standard model and beyond,
  Phys.\ Rev.\ D {\bf 66}, 014009 (2002)  [hep-ph/0112235]. 
\end{thebibliography}
\end{document}